\documentclass[fleqn,usenatbib]{mnras}

\usepackage{newtxtext,newtxmath}
\usepackage[T1]{fontenc}

\DeclareRobustCommand{\VAN}[3]{#2}
\let\VANthebibliography\thebibliography
\def\thebibliography{\DeclareRobustCommand{\VAN}[3]{##3}\VANthebibliography}

\usepackage{graphicx}	
\usepackage{amsmath}	
\usepackage{cleveref}
\usepackage{multirow}
\usepackage{tikz}
\usepackage{comment}

\newcommand{\tikzmark}[1]{\tikz[baseline,remember picture] \coordinate (#1) {};}

\tikzset{
  square arrow/.style={
    to path={-- ++(-10,-.25) -| (\tikztotarget)}
  }
}

\title[S2 orbital precession STVG]{Orbital precession of the S2 star in Scalar-Tensor-Vector-Gravity}

\author[R. Della Monica et al.]{
Riccardo Della Monica,$^{1}$\thanks{E-mail: rdellamonica@usal.es}
Ivan de Martino,$^{1}$\thanks{E-mail: ivan.demartino@usal.es}
Mariafelicia de Laurentis,$^{2,3}$\thanks{E-mail: mariafelicia.delaurentis@unina.it}
\\
$^{1}$Universidad de Salamanca, Departamento de Fisica Fundamental, P. de la Merced, Salamanca, ES\\
$^{2}$Dipartimento di Fisica, Universit\'a
di Napoli {}``Federico II'', Compl. Univ. di
Monte S. Angelo, Edificio G, Via Cinthia, I-80126, Napoli, Italy\\
$^{3}$ INFN Sezione  di Napoli, Compl. Univ. di
Monte S. Angelo, Edificio G, Via Cinthia, I-80126, Napoli, Italy
}

\date{Accepted XXX. Received YYY; in original form ZZZ}

\pubyear{2021}

\begin{document}
\label{firstpage}
\pagerange{\pageref{firstpage}--\pageref{lastpage}}
\maketitle

\begin{abstract}
We have obtained the first constraint of the parameter space of Scalar-Tensor-Vector-Gravity using the motion of the S2-star around the supermassive black hole at the centre of the Milky Way, and we did not find any serious tension with General Relativity. We used the Schwarzschild-like metric of Scalar-Tensor-Vector-Gravity to predict the orbital motion of S2-star, and to compare it with the publicly available astrometric data, which include 145 measurements of the positions,  44 measurements of the radial velocities of S2-star along its orbit, and the recent measurement of the orbital precession. We employed a Monte Carlo Markov Chain algorithm to explore the parameter space, and constrained the only one additional parameter of Scalar-Tensor-Vector-Gravity to $\alpha \lesssim 0.410$ at 99,7$\%$ confidence level, where $\alpha=0$ reduces this modified theory of gravity to General Relativity.
\end{abstract}




\section{Introduction}

The Scalar-Tensor-Vector-Gravity (STVG), also referred to as MOdified Gravity (MOG), is a generally covariant alternative theory of gravity developed by \citet{moffat_svtg} whose main aim is to resolve the puzzling mystery of dark matter  \citep{deMartinoChakrabarty2020}. In STVG, the new action depends not only on the Ricci scalar, $R$, describing the curvature of space-time, but also on a massive vector field $\phi^\alpha$, and treats as dynamical scalar fields both Newton's gravitational constant $G$ (whose Newtonian value is hereafter denoted as $G_N$) and the mass $\mu$ of the vector field. Massive test particles interact  with such a vector field, whereby a repulsive term appears in the geodesic equations affecting the trajectories on short scales ({\em i.e.} galactic, sub-galactic, or smaller scales) and balancing the  enhancement of the strength of the gravitational field due to the scalar fields.

STVG has been widely tested in its weak field limit \citep{moffat_galaxyrot1}. Indeed, by means of the Yukawa-like correction to the gravitational potential arising in the post-Newtonian approximation, STVG can explain rotation curves of spiral galaxies, multi-wavelengths observations of galaxy clusters, gravitational lensing effect, and cosmological datasets without resorting to dark matter (\citet{Brownstein2006,Moffat2011,moffat_galaxyrot1,moffat_clusters2,DeMartino2017}). Interestingly, some inconsistencies appears at the scale of dwarf galaxies (\citet{Haghi2016,deMartino2020}) shedding doubts on its effectiveness. At shorter scale, STVG was used to explain the mass-radius relation of white dwarfs and neutron stars, and the emission of gravitational waves from the coalescence of binary black holes (\citet{Moffat2016,lopez-romero,Banerjee2017}). While, on even smaller scales, STVG has not yet been sufficiently tested although Kerr-like black hole metrics have been derived \citep{moffat_blackhole1}. These results open a different avenue to test STVG in a strong field regime rather than a weak field limit. For instance, accretion disk around supermassive black holes (SMBH) and relativistic jets have been investigated without highlighting departures from General Relativity (GR) (\citet{Perez2017,Lopez_2017}). Using the results in (\citet{moffat_blackhole1}), \citet{rdm} investigated how the shadow of a black hole changes based on the spin of the black hole and the enhancement of the gravitational field due to the presence of the scalar field.  Additionally, the study of the orbital motion of a star around a black hole, based on the solution of the geodesic equations, may also lead to interesting results. For example, the astrometric motion of S2-star around the SMBH in the centre of the Milky Way, and the measurement of the orbital precession has been recently used to obtain interesting constraints on the parameter of the Yukawa-like gravitational potential arising in $f(R)$-gravity \citep{deMartino2021}. 

The SMBH in the centre of our own Milky Way \citep{Genzel2010} was firstly detected through radio observations as a point source named  Sagittarius A* (Sgr A*). Stars orbiting around Sgr A*  have been detected and monitored through the last three decades \citep{Genzel2010}. They move with large velocities ($\simeq 1000\,km/s$) in Keplerian orbits  pointing out that in the centre of the Galaxy must reside a  compact object of  mass of $M\simeq 4\times10^6M_\odot$ concentrated within a few hundreds Schwarzschild radii.  Observations of such a radio and X-ray emitting region  will be very useful to unveil signatures of modifications of gravity.  Indeed, several seminal analyses have been already carried out. For instance, studies of a SMBH surrounded by scalar field dark matter \citep{Hui2019}, analyses of the no-hair theorem with Srg A* \citep{Psaltis2016}, and fitting of the orbital motion of S2 star in different theories (\citet{Hees2017, deMartino2021}), among others. 

Here, we use publicly available astrometric and spectroscopic data of S2-star that have been collected during the past thirty years \citep{gillessen}, and the additional measurement of the orbital precession achieved by \citet{gravity}, to constrain the Schwarzschild-like metric of STVG. Our approach is based on the numerical integration of the geodesic equations and the exploration of the parameter space throughout a Monte Carlo Markov Chain (MCMC) analysis. In Section \ref{sec:stvg_background}, we briefly introduce the main aspect of the STVG. We describe the action, the field equations, and the corresponding Schwarzschild-like metric. We then explain the details of the geodesic equations which will be our main tool. In Section \ref{sec:data}, we briefly illustrate the data used in our analysis. In Section \ref{sec:methodology}, we explain the methodology used to carry out our statistical analysis. We give a detailed description of the model and  the priors employed in our pipelines. In Section \ref{sec:results}, we illustrate and discuss our results and, finally, in Section \ref{sec:conclusions} we give our final conclusions.

\section{Scalar-Tensor-Vector-Gravity}
\label{sec:stvg_background}

In STVG, the generally covariant action takes the form \citep{moffat_svtg}:
\begin{equation}
    \mathcal{S} = \mathcal{S}_{GR}+\mathcal{S}_M+\mathcal{S}_\phi+\mathcal{S}_S.
    \label{eq:stvg-action}
\end{equation}
Here the first term is the classical Hilbert-Einstein action and the second one is related to the ordinary matter energy-momentum tensor,
\begin{align}
    \mathcal{S}_{GR} &= \frac{1}{16\pi}\int d^4x\sqrt{-g}\frac{1}{G}R,&
    T^M_{\alpha\beta} &= -\frac{2}{\sqrt{-g}}\frac{\delta\mathcal{S}_M}{\delta g^{\alpha\beta}}.
    \label{eq:hilbert-einstein}
\end{align}
where $g$ denotes the determinant of the metric tensor $g_{\alpha\beta}$ and $R$ represents the Ricci scalar. On the other hand, the two additional terms, $\mathcal{S}_\phi$ and $\mathcal{S}_S$, encode the novel features of STVG:
\begin{align}
    \mathcal{S}_\phi =& -\int d^4x\sqrt{-g}\left(\frac{1}{4}B^{\alpha\beta}B_{\alpha\beta}-\frac{1}{2}\mu^2\phi^\alpha\phi_\alpha+V(\phi)\right),\\
    \mathcal{S}_S =& \int d^4x\sqrt{-g}\frac{\omega_M}{G^3}\left(\frac{1}{2}g^{\alpha\beta}\nabla_\alpha G\nabla_\beta G-V(G)\right)+\nonumber\\
    &+\int d^4x\frac{1}{\mu^2G}\left(\frac{1}{2}g^{\alpha\beta}\nabla_\alpha\mu\nabla_\beta\mu-V(\mu) \right),
\end{align}
where $\nabla_\alpha$ is the covariant derivative related to the metric tensor $g_{\alpha\beta}$, $\omega_M$ is a constant, $B_{\alpha\beta}$ is a Faraday tensor associated to the vector field $\phi_\alpha$ as follows
\begin{equation}
    B_{\alpha\beta}=\nabla_\alpha\phi_\beta-\nabla_\beta\phi_\alpha \,,  
\end{equation}
and $V(\phi)$, $V(G)$ and $V(\mu)$ are scalar potentials arising from the self-interaction associated with the vector field and the scalar fields, respectively. 

Upon minimization of the action in Eq. \eqref{eq:stvg-action}, the field equations in vacuum ({\i.e.} $T_{\alpha\beta}^M = 0$) read \citep{Moffat_2021}
\begin{align}
    G_{\alpha\beta}=&-\frac{\omega_M}{\chi^2}\biggl(\nabla_\alpha\chi\nabla_\beta\chi -\frac{1}{2}g_{\alpha\beta}\nabla^\sigma\chi\nabla_\sigma\chi\biggr)+
\\&-\frac{1}{\chi}(\nabla_\alpha\chi\nabla_\beta\chi-g_{\alpha\beta}\Box\chi)+\frac{8\pi}{\chi}T^\phi_{\alpha\beta},
\end{align}
where the scalar field $\chi = 1/G$, and $T^\phi_{\alpha\beta}$ is the gravitational $\phi$-field energy momentum tensor given by
\begin{equation}
    T^\phi_{\alpha\beta} = -\left({B_\alpha}^\sigma B_{\sigma\beta}-\frac{1}{4}g_{\alpha\beta}B^{\sigma\rho}B_{\sigma\rho}\right).
\end{equation}
Finally, assuming a spherically symmetric metric tensor,  a Schwarzschild-like black hole solution can be obtained \citep{moffat_blackhole1}:
\begin{align}
    ds^2=&\frac{\Delta}{r^2}dt^2-\frac{r^2}{\Delta}dr^2-r^2d\Omega^2\,,
    \label{eq:stvg-sch-metric}
\end{align}
where 
\begin{align}
    &\Delta=r^2-2G_NMr+\alpha G_NM\bigl((1+\alpha)G_NM-2r\bigr),\\
    &d\Omega^2=d\theta^2+\sin^2\theta d\phi^2.
\end{align}
Here,  the speed of light in vacuum is set to $c = 1$. Let us note that setting $\alpha=0$ returns the Schwarzschild metric. The result in Eq. \eqref{eq:stvg-sch-metric} is based on few additional hypothesis: {\em (i)} the gravitational coupling $G$ is constant, $\partial_\nu G = 0$ \citep{moffat_blackhole1, Moffat_2021}, and its value is $G=G_N(1+\alpha)$ where $\alpha$ is a free dimensionless parameter, whose particular value depends on the mass of the field source \citep{moffat_svtg}, and must be constrained through the observations;
{\em (ii)} The mass $\mu$ of the vector field $\phi^\alpha$ produces effects on kpc scales from the source \citep{moffat_galaxyrot1, moffat_galaxyrot2, moffat_clusters3} and can thus be neglected when solving field equations for compact objects; {\em (iii)} the parameter $\alpha$ also defines the proportionality constant between the mass and the fifth-force charge of a body by
    \begin{equation}
        \kappa = \sqrt{\alpha G_N}.
        \label{eq:kappa}
    \end{equation}
This novel interaction between massive bodies affects the motion of test particles as we will describe in \cref{sec:motion_test_particles}.

The metric element in Eq. \eqref{eq:stvg-sch-metric} represents the most general static spherically symmetric solution in STVG and it describes exactly the gravitational field produced by a non rotating point mass $M$ (and thus a fifth-force charge $Q = \sqrt{\alpha G_N} M$) in its neighborhood. We can, hence, use Eq. \eqref{eq:stvg-sch-metric} to describe the space-time around a gravitationally collapsed body (like a SMBH). As for the Reissner–Nordström \citep{nordstrom, reissner} and the Kerr solutions \citep{kerr}, the STVG black hole admits two horizons:
\begin{equation}
    r_{\pm} = G_NM\bigl(1+\alpha\pm\sqrt{1+\alpha}\bigr)\,,
    \label{eq:event_horizon}
\end{equation}
given by the roots of the $\Delta$ term in Eq. \eqref{eq:stvg-sch-metric}. For $\alpha = 0$ one recovers the Schwarzschild radius ($r_+ = r_- = 2G_NM$). Finally, due to the signs in Eq. \eqref{eq:event_horizon} the outer event horizon $r_+$ is always larger than $2G_NM$ when $\alpha > 0$ affecting both the dynamics of test particles and the shadow of a black hole in STVG (as shown in \citet{rdm}).

\subsection{Motion of test particles in STVG}
\label{sec:motion_test_particles}

In STVG,  the vector field gives rise to a modified version of the geodesic equations \citep{moffat_eqmotion}:
\begin{align}
    \left(\frac{d^2x^\alpha}{d\lambda^2}+\Gamma^\alpha_{\beta\rho}\frac{dx^\beta}{d\lambda}\frac{dx^\rho}{d\lambda} \right)=\frac{q}{m}{B^{\alpha}}_\beta\frac{dx^\beta}{d\lambda}.
    \label{eq:geodesic-equations}
\end{align}

While the left-hand side represents the standard terms of the geodesic equations,  on the right-hand side an extra force appears, usually referred to as fifth force \citep{moffat_svtg, moffat_galaxyrot1, moffat_eqmotion}, which behaves like a Lorentz-type force depending on the four-velocity of the particle and on a coupling constant, $q$, of the test particle with the vector field $\phi^\alpha$. This term is postulated by \cite{moffat_svtg} to be repulsive ($q>0$) in order to admit physically stable self-gravitating systems \citep{moffat_galaxyrot1}. Furthermore, in order to recover Einstein's Equivalence Principle the value of $q$ is assumed to be proportional to the mass of the particle itself, $q = \kappa m$ where $\kappa$ is given by Eq. \eqref{eq:kappa} \citep{moffat_eqmotion}.

In the case of a  black hole solution,  the vector field can be expressed in the 1-form coordinate basis associated to Eq. \eqref{eq:stvg-sch-metric} as (\citet{lopez-romero, rdm})
\begin{equation}
    \phi_\alpha = \left(-\frac{\sqrt{\alpha G_N}M}{r}, 0, 0, 0\right)
    \label{eq:vector_field},
\end{equation}
giving rise to a radial repulsive force similarly to that generated by an electric field of a point charge in electrodynamics.

Two distinct features of STVG, hence, both encoded in the parameter $\alpha$, have an influence on the motion of massive test particles. The enhanced gravitational constant increases the attractive effect of gravity on test particles, while the repulsive effect of the vector field counteracts this increased attraction. As shown by \cite{rdm}, a distinctive trait of test particles trajectories around a black hole in STVG is an increased value of the periastron advance, which explicitly depends on the parameter $\alpha$,  whose first-order analytical expression is given by
\begin{equation}
    \Delta\phi = \Delta\phi_{\rm GR}\left(1+\frac{5}{6}\alpha\right).
    \label{eq:stvg_precession}
\end{equation}
Here, $\Delta\phi_{\rm GR}$ is the usual expression of the orbital precession in GR,
\begin{equation}
    \Delta\phi_{\rm GR} = \frac{6\pi G_NM}{ac^2(1-e^2)}\,,
    \label{eq:gr_precession}
\end{equation}
and $a$ and $e$ are the orbital semi-major axis and eccentricity, respectively.

\section{Data}
\label{sec:data}
Our analysis uses publicly available astrometric and spectroscopic data that have been collected during the past thirty years. Specifically, we use:

\begin{itemize}
	\item 145 astrometric positions spanning a period from 1992.225 to 2016.38 that were retrieved from Table 5 in \citet{gillessen}. These data come from different observatories and instruments, more precisely: (a) data from 1992.224 to 2002 were collected using the speckle camera SHARP at the ESO New Technology Telescope  \cite[NTT, ][]{hoffman}, with uncertainties on astrometric position of order 3.8 mas; (b) measurements going from 2002 to 2016.38 were made using the Very Large Telescope (VLT) Adaptive Optics (AO) assisted infrared camera NAOS+CONICA instrument, better known as NACO \citep{lenzen, rousset}, and come with a rms uncertainty of order $\approx 400 \mu$as. 
	\item 44 spectroscopic measurements of the Brackett-$\gamma$ line by which the radial velocity of S2 were estimated \citep{gillessen}. The data collected before 2003 come from the instrument NIRC2, an AO imager and spectrometer located at the Keck observatory \citep{Ghez_2003}; data acquired after 2002, on the other hand, were collected with the Spectrograph for INtegral Field Observations in the Near Infrared (SINFONI), an AO-assisted integral field spectrometer at the VLT \citep{eisenhauer2003, bonnet}. 
\end{itemize}

Those data will serve to constrain STVG to fit the orbital motion  of S2-star.

\section{Modelling the orbital motion of  S2 in STVG}\label{sec:methodology}

The S2 star in the nuclear star cluster of the Milky Way orbits, with a period of $\sim 16$ years, the compact radio source Sgr A* which is the nearest SMBH candidate (\citet{Ghez_1998, Schodel_2002}). S2 has been extensively observed during the last three decades. At the pericentre (reached in 2002 and 2018), which is $\sim120$ AU (roughly 1400 gravitational radii) far way from Sgr A*, S2 reaches an orbital speed of $\sim 7700$ km/s. Thanks to its proximity to the SMBH and its high velocity ($\sim2.5\%c$) at the pericentre, astrometric and spectroscopic data  have been used to place constraints on the first post-Newtonian order of GR (\citet{do, gravity2018, gravity}). In particular, \citet{gravity} measured, for the first time,  the orbital precession of S2, and constrained its departure from the one predicted in GR as an extra parameter $f_{\rm SP}$: 
\begin{equation}
\Delta\phi = f_{\rm SP}\Delta\phi_{\rm GR}\,,
    \label{eq:gravity_precession}
\end{equation}
whose best fit value is $f_{\rm SP, obs} = 1.10\pm0.19$.
Here a value $f_{\rm SP}=0$ recovers Newtonian's gravity and $f_{\rm SP} = 1$ is consistent with GR. Equations \eqref{eq:stvg_precession} and \eqref{eq:gravity_precession} show how, already at the present stage, we could use the measurement of the orbital precession of S2 from \citet{gravity} to place constraints on the parameter $\alpha$ of the STVG theory. However, due to the presence in Eq. \eqref{eq:gr_precession} of the mass $M$ of the central source and of the orbital parameters $a$ and $e$ (which, on turn, depend on the particular orbital model and on the observed sky-projected positions of the star), one must retrieve them among the parameters of the orbital model in STVG to be fitted with   of the observational data in order to constrain $\alpha$. 

\subsection{Orbital model}
\label{sec:orbital_model}
In our model, the orbit for S2 is fully determined once the following 15 parameters are specified: 
\begin{gather}
    (\underbrace{M_\bullet,R_\bullet}_{\tikzmark{a}},\underbrace{T,t_p,a,e,i,\Omega,\omega}_{\tikzmark{b}},\underbrace{x_0,v_{x,0},y_0,v_{y,0},v_{\rm LSR}}_{\tikzmark{c}},\underset{\tikzmark{d}}{\alpha})\,.\label{eq:mcmc_parameters}\\
    \nonumber
  \tikz[remember picture]{\node[align=center, anchor = north](A){\baselineskip=2pt Central \\ object};}\quad
  \tikz[remember picture]{\node[align=center](B){Keplerian elements};}\quad
  \tikz[remember picture]{\node[align=center](C){Reference frame};}\quad
  \tikz[remember picture]{\node[align=center](D){STVG};}\quad
  \tikz[remember picture,overlay]{
    \draw[->] (a.south)++(0,1.5ex) to (A.north) ;
    \draw[->] (b.south)++(0,1.5ex) to (B.north) ;
    \draw[->] (c.south)++(0,1.5ex) to (C.north) ;
    \draw[->] (d.south)++(0,0) to (D.north) ;
  }
\end{gather}  
The first two parameters, $M_\bullet$ and $R_\bullet$, denote the mass and the distance from Earth of the central SMBH around which S2 revolves. Next, we have the seven Keplerian elements \citep{green1985spherical},  the five reference-frame parameters $(x_0,v_{x,0},y_0,v_{y,0},v_{\rm LSR})$, and finally the parameter $\alpha$ of the STVG theory. 

While all the astrometric data have been corrected in order to be referred to the ‘Galactic Centre infrared reference system’ (we refer to \citet{Plewa} for an extensive analysis), the reference frame can still be affected by a zero-point offset and drift \citep{gillessen}. To account for these effects, four reference-frame parameters must be considered $(x_0,v_{x,0},y_0,v_{y,0})$, which are currently regarded as limiting factor in the precision of the determination of the orbital precession \citep{gravity}. The last one, $v_{\rm LSR}$, is introduced to account for systematic effects when radial velocities are corrected for the local standard of rest (LSR) \citep{gillessen}.

Importantly, the seven Keplerian elements play a crucial role. They are the orbital period $T$, the time of the pericentre passage $t_p$, the semi-major axis $a$, the eccentricity $e$, the inclination $i$, the angle of the line of nodes $\Omega$, and the angle from ascending node to pericentre $\omega$, respectively. The first four parameters describe the in-orbital-plane motion of the star, while the last three serve to project the orbit in the reference frame of a distant observer. Since a relativistic orbit is known to depart from a purely Keplerian orbit, these parameters are time-dependent. In particular, the orbital precession effect is described by a time variation of the parameter $\omega$ \citep{poisson_will_2014}. For this reason, the Keplerian ellipse associated with these parameters must be regarded as the one osculating the true orbit at a certain time. In our analysis this set of parameters is important because it is strictly related to the initial conditions from which we start our numerical integration of the geodesics. More specifically, we aim at integrating numerically the four second-order differential equations in Eq. \eqref{eq:geodesic-equations} for the unknown functions $\{t(\lambda), r(\lambda), \theta(\lambda), \phi(\lambda)\}$, where $\lambda$ represents an affine parameter that in our case coincides with the proper time of the star. Such equations need second-order initial data at $\lambda = 0$, namely the four positions of the star at the initial time and values for the components of its four-velocity. Due to the spherical symmetry of the metric in Eq. \eqref{eq:stvg-sch-metric}, one can always set (by means of a rotation of the reference frame) $\theta(0) = \pi/2$ and $\dot{\theta}(0)=0$, meaning that the star and its spatial velocity initially lie on the equatorial plane of the reference system. These two conditions yield  $\ddot{\theta}(\lambda) = 0$ identically, constraining the particle to a planar motion on the equatorial plane.
To set initial conditions on $r$ and $\phi$, one can resort to the orbital elements. For the sake of simplicity, we assume that $\lambda = 0$ coincides with the time of the apocentre passage $t_a = t_p-T/2$.  Under these assumptions we can set $\dot{r}(0) = 0$ and $\phi(0)=\Phi(0)=\pi$ where $\phi$ and $\Phi$ are the true anomaly and the eccentric anomaly, respectively. Thus, the Cartesian coordinates of the star, expressed in the orbital plane, at the initial time are:
\begin{align}
    (x_{\rm orb}, y_{\rm orb}) &= \left(a(\cos\Phi-e),\, a\sqrt{1-e^2}\sin\Phi\right)=\nonumber\\&=(-a(1+e),0),
\end{align}
from which we retrieve the value for $r(0)=a(1+e)$. The Cartesian components of the velocity in the equatorial plane, on the other hand, read
\begin{align}
    (v_{\rm x, orb}, v_{\rm y,orb}) &= \left(-\frac{2\pi}{T}\frac{a^2}{r}\sin\Phi,\, \frac{2\pi}{T}\frac{a^2}{r}\sqrt{1-e^2}\cos\Phi\right)=\nonumber\\
    &=\left(0,\, \frac{2\pi}{T}\frac{a^2}{r}\sqrt{1-e^2}\right).
\end{align}
and yield a value for $\dot{\phi}(0)=\frac{2\pi}{T}(1-e)^{1/2}(1+e)^{-3/2}$. Finally, the initial condition  $\dot{t}(0)$ is obtained upon applying the normalization condition to the four-velocity vector, {\em i.e.} $g_{\alpha\beta}u^\alpha u^\beta = -c^2$.\par
Starting from those initial conditions a numerical integration of Eqs. \eqref{eq:geodesic-equations} is carried out by means of a Runge-Kutta 5(4) algorithm \citep{dormand_prince}. In order to compare the numerically integrated orbit for S2 (in the SMBH reference frame) with astrometric and radial velocity data (in the observer reference frame), a projection must be performed by means of the Thiele-Innes elements \citep{thieleinnes}:
\begin{align}
    & \mathcal{A}=\cos\Omega\cos\omega-\sin\Omega \sin\omega \cos i,\\
    & \mathcal{B}=\sin\Omega \cos\omega+\cos\Omega \sin\omega \cos i\,,\\
	& \mathcal{C}=-\sin\omega \sin i,\\
    & \mathcal{F}=-\cos \Omega \sin\omega-\sin\Omega \cos\omega \cos i,\\
    & \mathcal{G}=-\sin\Omega\sin\omega+\cos\Omega \cos\omega \cos i,\\
    & \mathcal{H}=-\cos\omega \sin i.
\end{align}
In particular, the position $(x,y)$ on the plane of the observer and the distance $z$ from it can be obtained with:
\begin{align}
     x&=\mathcal{B}x_{\rm orb}+\mathcal{G}y_{\rm orb},\\
     y&=\mathcal{A}x_{\rm orb}+\mathcal{F}y_{\rm orb},\\
     z&=\mathcal{C}x_{\rm orb}+\mathcal{H}y_{\rm orb} + R_\bullet\,\label{eq:z_obs}.
\end{align}
Velocities, on the other hand, read
\begin{align}
 	v_x &= \mathcal{B}v_{\rm x, orb}+\mathcal{G}v_{\rm y, orb},\\
 	v_y &= \mathcal{A}v_{\rm x, orb}+\mathcal{F}v_{\rm y, orb},\\
	v_z &= -(\mathcal{C}v_{\rm x, orb}+\mathcal{H}v_{\rm y, orb})\,.
\end{align}
The minus sign in $v_z$ represents a positive radial velocity during the approaching phase and a negative one during recession.\par
Finally, in order to obtain quantities that can be comparable to the observed ones, we need to take into account some classical and relativistic effects on the orbit. 

\subsubsection{The Rømer delay}
Due to the inclination of the orbit with respect to the observer's sky, the star is not always at the same distance from the observer. This implies that the time of arrival of a photon from the S2 star to the Earth is modulated by a factor depending on the distance between the points of emission and observation. The classical expression of this effect, called Rømer delay, is given by:
\begin{equation}
    t_{\rm em} = t_{\rm obs}+\frac{z(t_{\rm em})}{c},
    \label{eq:roemerdelay1}
\end{equation}
where $t_{\rm em}$ is the unknown emission time, $z(t_{\rm em})$ is the distance of the star from the observer (see Eq. \eqref{eq:z_obs}) at the time of emission and $t_{\rm obs}$ is the observation date. Due to the presence of $t_{\rm em}$ on both sides of Eq. \eqref{eq:roemerdelay1}, in order to solve for $t_{\rm em}$ one can either use an iterative method \citep{grould} or approximate the expression with
\begin{equation}
	t_{\rm em} = t_{\rm obs} - \frac{z(t_{\rm em})}{c}\approx t_{\rm obs}-\frac{z(t_{\rm obs})}{c}\left(1-\frac{v_z(t_{\rm obs})}{c}\right)\,,
	\label{eq:roemerdelay2}
\end{equation}
as shown by \citet{gravity2018}, which directly allows to compute the emission time from the observation one. This first-order approximation is sufficient for our purposes, since higher order terms would lead to corrections on the order of minutes  which are negligible at the current level of accuracy (see for example the Supplementary Materials of \citet{do}). The resulting modulation on the arrival time of photons ranges from a minimum of $\sim-5$ days at pericentre to a maximum of $\sim7.5$ days at apocentre and affects both the sky-projected position of the star and its observed radial velocity.

\subsubsection{Frequency shift}
Spectroscopic observations of the S2 star allow to detect a frequency shift $\zeta$ that can be related to the radial velocity of the star as measured by a distant observer:
\begin{equation}
	\zeta=\frac{\Delta \nu}{\nu} = \frac{\nu_{\rm em}-\nu_{\rm obs}}{\nu_{\rm obs}}= \frac{v_z}{c}\,.
\end{equation}
A fraction of this frequency shift is ascribable to the actual orbital motion of the star, through the special relativistic (longitudinal and transverse) Doppler effect, $\zeta_D$,
\begin{equation}
    \zeta_D=\frac{\sqrt{1-\frac{v^2(t_{\rm em})}{c^2}}}{1-\vec{k}\cdot\vec{v}(t_{\rm em})},
\end{equation}
where $\vec{v}$ is the spatial velocity of the star, and $\vec{k}$ is the spatial part of the wave vector of the observed light ray, so that the scalar product $\vec{k}\cdot\vec{v} = v_z/ c$ represents the projection of $\vec{v}$ along the line of sight.

A significant fraction of the frequency shift, $\zeta_{G}$, on the other hand, is expected to be related to the gravitational redshift produced by the space-time curvature near the central SMBH. In particular, assuming that the observer is in a region where the space-time curvature is negligible (which is applicable to an infinitely distant observer in an asymptotically flat space-time), we have
\begin{equation}
    \zeta_G = \frac{1}{\sqrt{|g_{00}(t_{\rm em},\vec{x}_{\rm em})|}},
\end{equation}
where $g_{00}$ is the coefficient of the time component of the metric in Eq. \eqref{eq:stvg-sch-metric}. Under these assumptions, the total redshift amounts to:
\begin{align}
	1+\zeta = \zeta_D\cdot\zeta_G\,,
\end{align}
Since all the quantities are evaluated at $t=t_{\rm em}$, the Rømer delay is naturally taken into account in our radial velocity simulation.

\subsubsection{Higher-order relativistic effects}
While, in principle, other higher-order effects (such as the Shapiro delay, the Lense-Thirring effect on both the orbit and the photon due to a possible rotation of the SMBH or the deflection of the light rays emitted by the star) could arise in the determination of the sky-projected positions and radial velocity of the star, \citet{grould} have shown that at the current stage only the Rømer delay and the periastron advance have a non negligible effect on observations.

\subsection{Methodology}\label{subsec:methods}

In order to constrain the parameter space shown in Eq. \eqref{eq:mcmc_parameters}, we use the data presented in Section \ref{sec:data}, and predicted the orbit of the S2 star as explained in Section \ref{sec:orbital_model}. Specifically, we use a MCMC algorithm implemented in \texttt{emcee}  \citep{emcee} to perform a Bayesian sampling of the posterior distribution of our parameters. We carry out two analysis: first, we adopt a {\em Conservative Approach} (CA) where we state that only small deviations from GR are expected.
Therefore, in the CA, we adopt Gaussian priors centered on the best fit values of the central mass, distance and Keplerian elements as given by \citet{gravity}. For the reference frame parameters, on the other hand, we adopt Gaussian priors from \cite{Plewa}. These priors are listed as second (central value of the parameters) and third (standard deviation) columns of Table \ref{tab:priors}. While the corresponding references are given in the fourth column. 

Finally, we also adopt a {\em Questioner Approach} (QA) stating that no knowledge {\em a priori} exists on the orbital model of the S2 star in STVG. Therefore, all parameters may in principle deviate from the GR ones because the gravitational field itself is different. Thus, we adopt flat priors on all model parameters as reported in the fifth column of Table \ref{tab:priors}, { except for the orbital period $T$ and pericentre passage $t_p$ for which we adopted the same Gaussian priors as in the CA. This particular choice is due to the fact that data from \cite{gravity}, which means data-points for the pericentre passage, are not publicly available and, hence, these parameters cannot be properly constrained}.

In both CA and QA analysis, since we have {\bf \em no a priori constraints}, we set a uniform prior on the parameter $\alpha$ of STVG to cover an appropriate range which we have sampled logarithmically. It is worth to remark that $\alpha$ scales with the mass of the gravitational source \citet{moffat_galaxyrot1} and, hence, we expect $\alpha\sim0.03$ for a gravitational mass of $\sim10^6M_\odot$ \citep{Moffat_2008}.
All the priors and the corresponding references are reported in Table \ref{tab:priors}.\par

\begin{table}
\setlength{\tabcolsep}{4pt}
\centering
\caption{The sets of priors used to carry out our MCMC analysis. Priors on the orbital parameters of S2 are the most updated ones from \citet{gravity} while those on the reference frame are taken from \citet{Plewa}. They are set as Gaussian priors and reported in the second and third columns. The forth column lists the corresponding references. While, the uniform priors used in the QA are reported in the last column. The interval in which $\alpha$ varies uniformly, for both the CA and QA analysis, has been set heuristically.}
\begin{tabular}{lllcr}
\hline
Parameter                   & $\mu$      & $\sigma$ & \multicolumn{1}{l}{Reference} & Uniform\\ \hline
$M_\bullet$ $(10^6M_\odot)$ & 4.261      & 0.012    & G  &    $[3,\, 5]$                        \\
$R_\bullet$ (kpc)           & 8.2467     & 0.0093   & G   &   $[7,\, 9]$                        \\ \hline
$T$ (yr)                    & 16.0455    & 0.0013   & G   &   $-$  
\\
$t_p$ (yr)                  & 2018.37800 & 0.00017  & G   &   $-$
\\
$a$ (mas)                   & 125.058    & 0.044    & G   &     $[115,\,135]$                      \\
$e$                         & 0.884649   & 0.000079 & G    &    $[0.87,\, 0.90]$                      \\
$i$  ($^\circ$)             & 134.567    & 0.033    & G   &     $[131,\, 137]$                      \\
$\omega$  ($^\circ$)        & 66.263     & 0.030    & G   &     $[62,\, 69]$                      \\
$\Omega$  ($^\circ$)        & 228.171    & 0.031    & G    &    $[224,\, 230]$                      \\ \hline
$x_0$ (mas)                 & -0.2       & 0.2      & P    &    $[-10,\, 10]$                      \\
$y_0$ (mas)                 & 0.1        & 0.2      & P    &    $[-10,\, 10]$                      \\
$v_{x,0}$ (mas/yr)          & 0.05       & 0.1      & P    &    $[-0.1,\, 0.1]$                     \\
$v_{y,0}$ (mas/yr)          & 0.06       & 0.1      & P    &    $[-0.1,\, 0.1]$                      \\
$v_{\rm LSR}$                   & -1.6       & 1.4      & G  &  $[-50,\, 50]$                          \\ \hline
$\alpha$                   & \multicolumn{4}{r}{[$0,2$]}                                    \\ \hline
\multicolumn{5}{l}{\small G: \cite{gravity}}\\
\multicolumn{5}{l}{\small P: \cite{Plewa}}
\label{tab:priors}
\end{tabular}
\end{table}

At each iteration, the Bayesian sampler draws random samples for the 15 parameters in Eq. \eqref{eq:mcmc_parameters}, for each of the 30 chains that we have initialized. These are then used to compute the orbit of the S2 star following the prescriptions of our orbital model in Section \ref{sec:orbital_model}. Finally, the log-likelihood of that particular set of parameters is assessed by comparison with the data as: 
\begin{equation}
	\log\mathcal{L} = \log\mathcal{L}_{\rm P} + \log\mathcal{L}_{\rm RV} + \log\mathcal{L}_{\Delta\phi}
	\label{eq:log-likelihood}
\end{equation}
where $\log\mathcal{L}_{\rm P}$ is the log-likelihood related to the sky-projected positions of S2,
\begin{equation}
	\log\mathcal{L}_{\rm P} = -\sum_i\frac{(x_{\rm obs}^i-x_{\rm th}^i)^2}{{2(\sigma_{x,{\rm obs}}^i})^2}-\sum_i\frac{(y_{\rm obs}^i-y_{\rm  th}^i)^2}{{2(\sigma_{y,{\rm obs}}^i})^2},
\end{equation}
$\log\mathcal{L}_{RV}$ is the likelihood related to  the radial velocity measurements,
\begin{equation}
	\log\mathcal{L}_{\rm RV} = -\sum_i\frac{(v_{\rm z, obs}^i-\textrm{v}_{\rm z, th}^i)^2}{{2(\sigma_{v_z,{\rm obs}}^i})^2}
\end{equation}
and $\log\mathcal{L}_{\Delta\phi}$ is the log-likelihood related to measurement of the orbital precession
\begin{equation}
	\log\mathcal{L}_{\Delta\phi} = -\frac{(f_{\rm SP, obs}-f_{\rm SP, th} )^2}{2\sigma_{f_{\rm SP, obs}}^2},
	\label{eq:log-likelihood-precession}
\end{equation}
where $f_{\rm SP, obs}$ and $\sigma_{f_{\rm SP, obs}}$ are the value and uncertainty given in Section \ref{sec:methodology}, while $f_{\rm SP, th}=\Delta\phi/\Delta\phi_{\rm GR}$, where $\Delta\phi$ and $\Delta\phi_{\rm GR}$ are the analytical expressions for the first-order orbital precession given in \eqref{eq:stvg_precession} and \eqref{eq:gr_precession}, respectively. Here, subscripts "obs" and "th" indicate observational and theoretical ({\em i.e.} predicted) quantities, respectively. 

Convergence of the sampling is assessed by computing the averaged autocorrelation time, $\tau$, of the Markov chains \citep{Goodman2010} and guaranteeing that the length $\ell$ of the chains is $\ell>\mathcal{K}\cdot\tau$ (where in our case $\mathcal{K}=100$) and that the estimates of $\tau$ does not vary (on average) more than 1$\%$ \citep{emcee}.

We always perform two runs of the MCMC. In the first one,  we set $\log\mathcal{L}_{\Delta\phi} = 0$ in our log-likelihood. This corresponds to considering only the data-points for the sky-projected positions and the radial velocity measurements in our fitting procedure. In a second run   we use, along with the same dataset, the data-point for the precession as measured by \citet{gravity}, by restoring the expression in Eq. \eqref{eq:log-likelihood-precession}.

\begin{figure*}
    \centering
    \includegraphics[width = \textwidth]{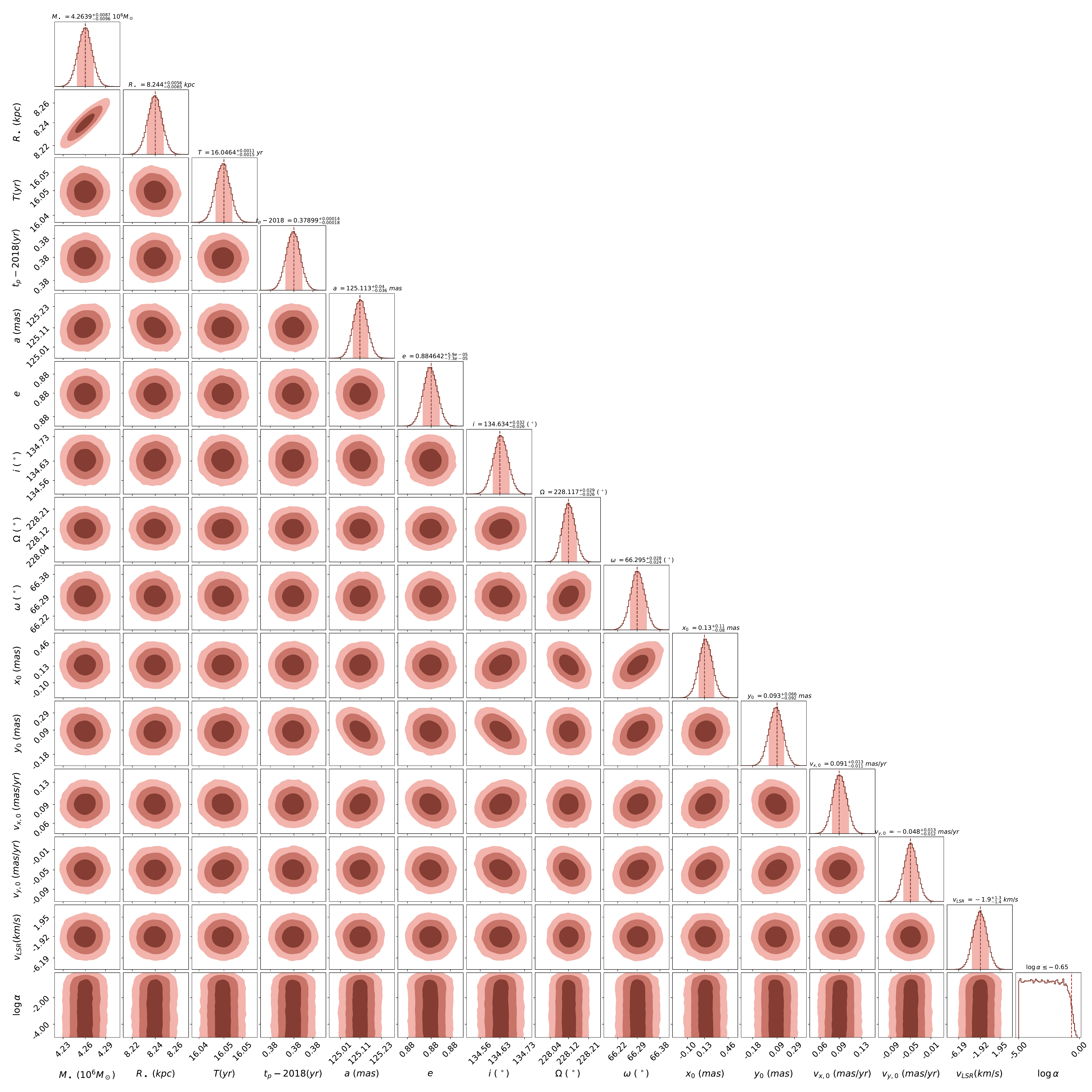}
    \caption{MCMC posterior distributions of the parameters of  the  orbital model of the S2 star in STVG in the CA case (taking into account the data-point for the precession), while in the histograms we report the 1$\sigma$ confidence level for all the parameters as a shaded area.}
    \label{fig:mcmc}
\end{figure*}

\section{Results and Discussions}\label{sec:results}

In this section we report and discuss the results of both CA and QA analyses. 
All the best fit parameters are listed in Table \ref{tab:results}, where the first column indicates parameters of our model, the second and third columns  report the best-fitting values from \citet{gillessen} and \citet{gravity}, respectively. Fourth and fifth columns list the median value and the 68\% confidence intervals that are computed estimating the percentiles of the posterior distributions  on our CA analysis. Similarly, the last column reports the results for our QA analysis. We also specify the approach used to obtain the mock orbits:  \citet{gillessen} integrate the orbits using a Keplerian approach,  \citet{gravity}, on the other hand, use a Post-Newtonian approach, while we integrate the geodesic equations. In the last two rows, we report the constraints on the STVG parameter, $\alpha$, at 99,7$\%$ confidence level and the mean value of the distribution, respectively.

\begin{table*}
    \centering
    \setlength{\tabcolsep}{12pt}
    \renewcommand{\arraystretch}{1.5}
    \caption{The complete set of values for the parameters of our model, resulting from the MCMC analysis. Columns 2 and 3 report the best-fitting values from \citet{gillessen} and \citet{gravity} for comparison. Columns 4 and 5 are the  results of the posterior analysis in the two cases considered (without and with precession measurements) in our CA analysis, while column 6 reports the results from the QA. In the third row, "Keplerian", "Post Newtonian", and "Geodesic" refer to the approach used in the analysis to obtain the orbits ({\em i.e.} which equations of motion are integrated).
    The last two rows report the constraints we have placed on the STVG parameter $\alpha$ at 99,7$\%$ confidence level and the mean value of the distribution, respectively.}
    \begin{tabular}{lcclcr}
        \hline
        Parameter & \citet{gillessen} & \citet{gravity} & \multicolumn{3}{c}{STVG} \\
         Model & Keplerian & Post Newtonian & \multicolumn{3}{c}{Geodesic}  \\ \cline{4-6}
         &           &                & $\underset{(\textrm{w/o precession})}{\rm CA}$    & $\underset{(\textrm{w/ precession})}{\rm CA}$ & $\underset{(\textrm{w/ precession})}{\rm QA}$ \\
        \hline
        $M_\bullet$ ($10^6M_\odot$) & $4.3\pm0.15$ & $4.261\pm0.012$ & $4.2632_{-0.0092}^{+0.0094}$ & $4.2639_{-0.0096}^{+0.0087}$ & $4.07_{-0.22}^{+0.20}$ \\
        $R_\bullet$ (kpc) & $8.17\pm0.15$ & $8.2467\pm0.0093$ & $8.2436_{-0.0062}^{+0.0059}$ & $8.244_{-0.0065}^{+0.0056}$ & $7.91\pm0.21$ \\
        $T$ (yr) & $16.0\pm0.02$ & $16.0455\pm0.0013$ & $16.046_{-0.0011}^{+0.0015}$ & $16.0464_{-0.0015}^{+0.0011}$ & $15.957_{-0.034}^{+0.028}$ \\
        $T_p - 2018$ & $0.37897\pm0.01$ & $0.379\pm0.00016$ & $0.379_{-0.00018}^{+0.00014}$ & $0.37899_{-0.00018}^{+0.00014}$ & $0.353_{-0.016}^{+0.015}$ \\
        $a$ (mas) & $125.5\pm0.9$ & $125.058\pm0.041$ & $125.113_{-0.033}^{+0.043}$ & $125.113_{-0.036}^{+0.040}$ & $128.1_{-1.2}^{+1.5}$ \\
        $e$ & $0.8839\pm0.0019$ & $0.884649\pm0.000066$ & $0.884639_{-0.000070}^{+0.000062}$ & $0.884642_{-0.000073}^{+0.000059}$ & $0.8893_{-0.0028}^{+0.0023}$ \\
        $i$ $(^\circ)$ & $134.18\pm0.4$ & $134.567\pm0.033$ & $134.635_{-0.029}^{+0.031}$ & $134.634_{-0.026}^{+0.032}$ & $133.42_{-0.55}^{+0.49}$ \\
        $\Omega$ $(^\circ)$ & $226.94\pm0.6$ & $228.171\pm0.031$ & $228.12\pm0.027$ & $228.117_{-0.026}^{+0.029}$ & $226.74_{-0.60}^{+0.78}$ \\
        $\omega$ $(^\circ)$ & $65.51\pm0.57$ & $66.263\pm0.031$ & $66.294_{-0.024}^{+0.028}$ & $66.295_{-0.024}^{+0.028}$ & $66.07_{-0.78}^{+0.59}$ \\
        $x_0$ (mas) & $-0.31\pm0.34$ & $-0.9\pm0.14$ & $-0.15_{-0.09}^{+0.10}$ & $-0.13_{-0.11}^{+0.08}$ & $-1.1_{-0.26}^{+0.19}$ \\
        $y_0$ (mas) & $-1.29\pm0.44$ & $0.07\pm0.12$ & $0.084_{-0.086}^{+0.073}$ & $0.093_{-0.092}^{+0.066}$ & $-2.22_{-0.66}^{+0.75}$ \\
        $v_{x,0}$ (mas/yr) & $0.078\pm0.037$ & $0.08\pm0.01$ & $0.093_{-0.013}^{+0.010}$ & $0.091_{-0.011}^{+0.013}$ & $0.117_{-0.046}^{+0.047}$ \\
        $v_{y,0}$ (mas/yr) & $0.126\pm0.047$ & $0.0341\pm0.0096$ & $0.049_{-0.014}^{+0.012}$ & $0.048_{-0.013}^{+0.012}$ & $-0.068_{-0.070}^{+0.077}$ \\
        $v_{\rm LSR}$ (km/s) & $8.9\pm4$ & $-1.6\pm1.4$ & $-1.9_{-1.5}^{+1.2}$ & $-1.9_{-1.4}^{+1.3}$ & $-16.1_{-7.5}^{+8.0}$ \\ \hline
        $\alpha$ (upper limit) & & & $\lesssim 0.637$ & $\lesssim 0.410$ & $\lesssim 0.548$ \\
        $\alpha$ (mean value) & & & $0.030$ & $0.027$ & 0.041 \\\hline
    \end{tabular}
    \label{tab:results}
\end{table*}

\subsection{Conservative Approach}

\begin{figure}
    \centering
    \includegraphics[width = \columnwidth]{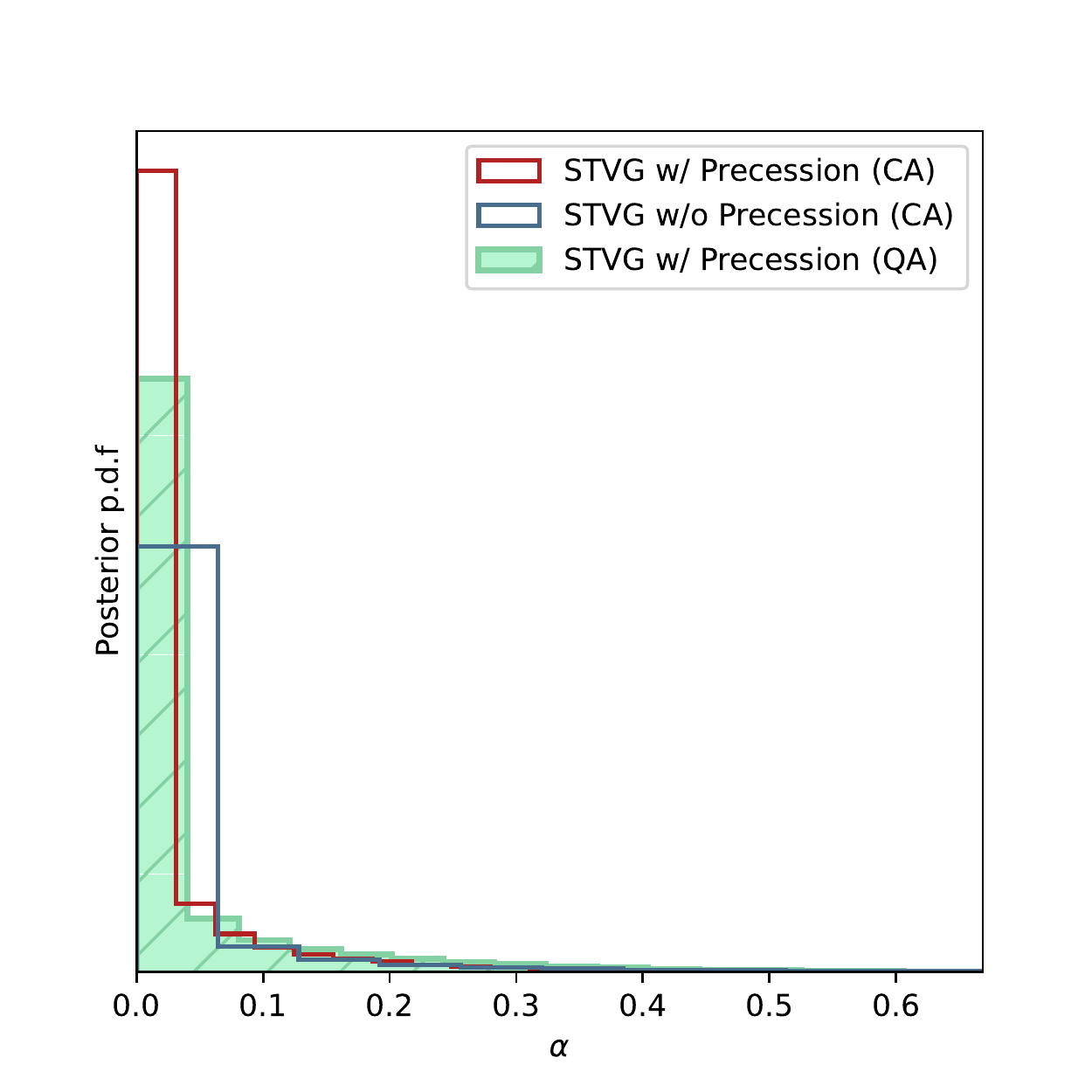}
    \caption{The figure depicts the posterior probability density function of the parameter $\alpha$ in linear space. The red and blue histograms shows the results with and without the precession data-point from \citet{gravity}, respectively, for the CA case.  The green shaded histogram is the posterior distribution of $\alpha$ including the information on the orbital precession in the QA analysis.}
    \label{fig:prevession_vs_noprecession}
\end{figure}

For our CA analysis, we run 30 MCMC chains with random starting points in the parameter space. We use Gaussian priors on all parameters, except for the STVG parameter. Gaussian priors on the SMBH and orbital parameters are centered on the best fits values obtained by \citet{gravity}, while the ones on the reference frame parameters are taken from \citet{Plewa}. Behind this choice, there is the conservative assumption that we do not expect to find large departures from GR. On the other hand, the priors on $\alpha$ are uniform reflecting the lack of previous constraints at those scales. In CA, we run two set of MCMC analysis: the first without including the measurement of the orbital precession and the second one including it. As it stands out from the results reported in the fourth and fifth columns of Table \ref{tab:results}, respectively, we do not find any relevant difference between the two MCMC runs in the median values and confidence intervals obtained from the posterior distributions. 
Therefore, in Figure \ref{fig:mcmc},  we report only the post burn-in posterior density distributions of the parameters for the MCMC run that includes the precession measurement.  
The value of the STVG parameter, $\alpha$, has a mean value of $0.030$ and $0.027$, and an upper limit at 99,7$\%$ confidence level of $\lesssim 0.637$ and $\lesssim 0.410$ for the two runs,  respectively. Adding the precession measurement reduces the upper bound on $\alpha$ of $\sim 35\%$. This is also visually reported in Figure \ref{fig:prevession_vs_noprecession}, where the posterior distributions (in linear space)  of the parameter $\alpha$ are shown  in blue and red for the two MCMC runs, respectively.

As expected, $\alpha$ has not a lower bound, and it agrees with GR at $1\sigma$. Up-to-date, this is the first constraint of STVG at scale of the SMBH on the center of our Galaxy. It also agrees with the theoretical estimation of $\alpha=0.03$ for a gravitational system whose mass is of the order of $\sim 10^6 M_\odot$ \citep{Moffat_2008}. 

\begin{figure*}
    \centering
    \includegraphics[width = \textwidth]{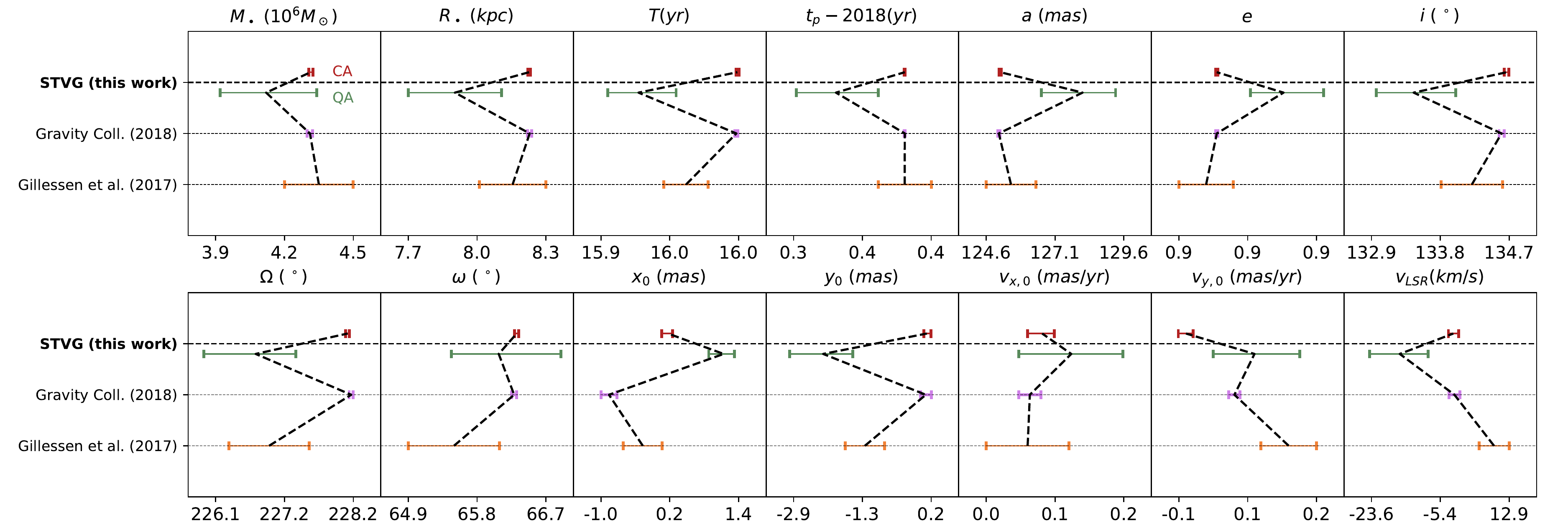}
     \caption{Comparison between the best-fitting values of the model parameters from \citet{gillessen} (third line), \citet{gravity} (second line),   and  the results of our posterior analysis (first line) for both the CA (red) and the QA (green). The amplitude of the intervals is considered at $1\sigma$ level.}
    \label{fig:gillessen_gravity_stvg}
\end{figure*}

 Finally, Figure \ref{fig:gillessen_gravity_stvg} shows visually the comparison between our constraints and the ones from \citet{gillessen} and  \citet{gravity} reported in the second and third columns of Table \ref{tab:results}, respectively.  As expected, the results of our CA approach, reported in red, agree with the ones from both analysis without tensions arising.

\subsection{Questioner Approach} 
\begin{figure*}
    \centering
    \includegraphics[width = \textwidth]{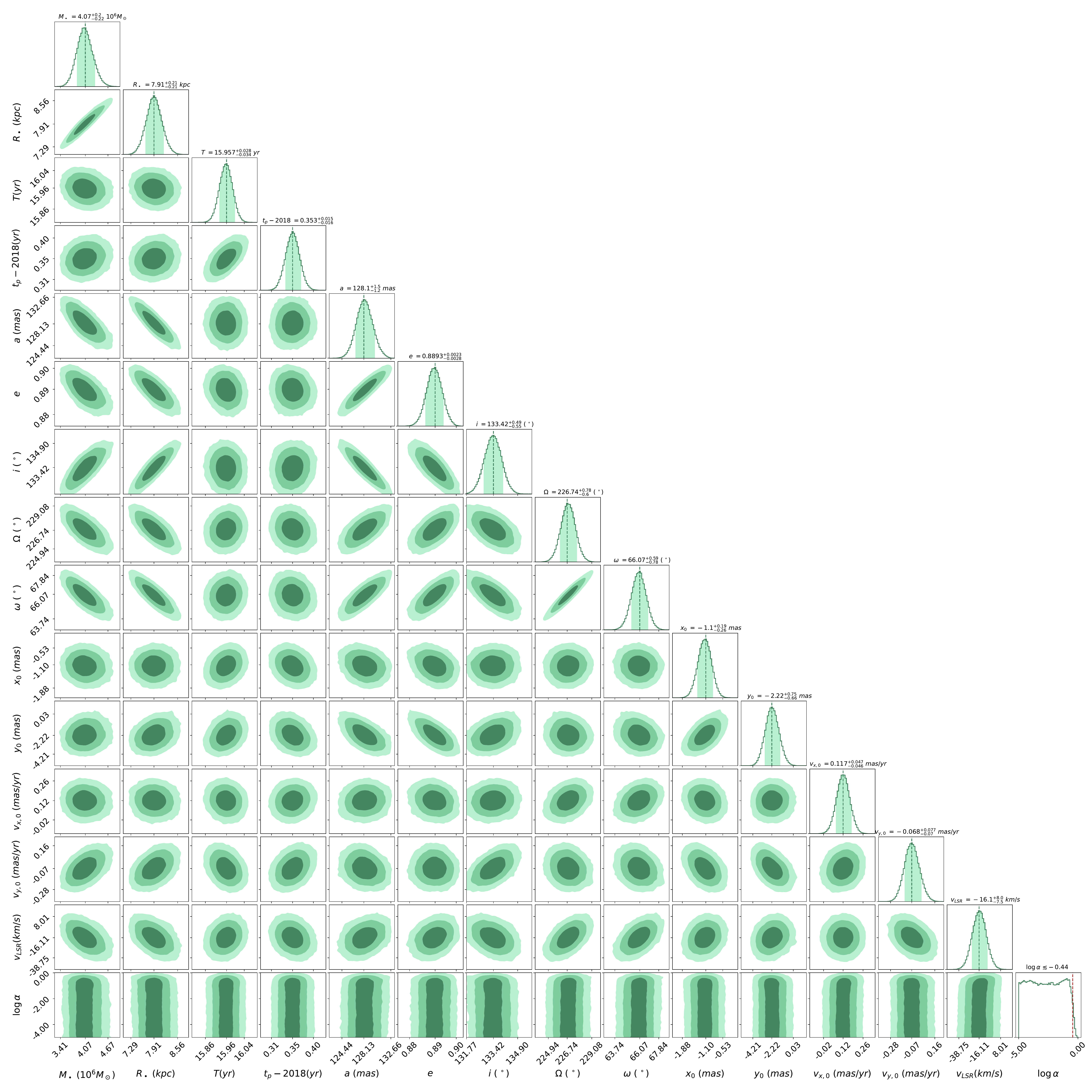}
    \caption{The same as Fig. \ref{fig:mcmc} particularized for the QA case.}
    \label{fig:mcmc_qa}
\end{figure*}

In our QA analysis, we run 30 MCMC chains with random starting points and flat priors on all parameters, except for the orbital period, and the time of the passage at the pericentre, as explained in Section \ref{subsec:methods}. Let us remark that this choice wants to reflect the lack of {\em a priori} studies of this gravitational system in STVG and, hence, one could expect some departures from GR arising also in other parameters. In QA, we run only one MCMC analysis including also the measurement of the orbital precession\footnote{We checked that results obtained without including the precession information are self-consistently weaker.}.  The median value and the corresponding 68\% confidence level of all parameters are reported in the sixth column of Table \ref{tab:results}, while the post burn-in posterior density distributions of the parameters 
is shown in Figure \ref{fig:mcmc_qa}. The model seems to predict a less massive black hole of mass $M_\bullet=4.07_{-0.22}^{+0.20}\times10^6 M_\odot$ which is also located at a shorter distance from the Earth, $R_\bullet=7.91\pm0.21$ kpc. This causes a slight change also in the orbital parameters. For instance, the semi-major axis and eccentricity are larger, while orbital period and inclination are shorter than ones in \citet{gillessen} and \citet{gravity}. This is understandable thinking to the fact that the gravitational field itself is enhanced as soon as $\alpha>0$. Nevertheless, all parameters are still compatible with ones from \citet{gillessen} and \citet{gravity} at most at 3$\sigma$, hence not really favouring a departure from GR. This is also shown in Figure \ref{fig:gillessen_gravity_stvg} where QA results are reported in green to be compared with other results listed in Table \ref{tab:results}. Finally, The value of the STVG parameter, $\alpha$, has a mean value of $0.041$ and an upper limit at 99,7$\%$ confidence level of $\lesssim 0.548$, while is posterior distribution is reported in Figure \ref{fig:prevession_vs_noprecession} as the green shaded histogram.

\section{Conclusions}\label{sec:conclusions}

The weak field limit of STVG has been widely  used to explain 
astrophysical phenomena at galactic, extra-galactic  and cosmological scales without resorting to dark matter (\citet{moffat_galaxyrot1,moffat_clusters2,Brownstein2006,Moffat2011,DeMartino2017}). Although, some inconsistencies exist at the scale of dwarf galaxies (\citet{Haghi2016,deMartino2020}), STVG can potentially offers a solution to the puzzling dark matter mystery. 

On stellar scales, STVG is not as well tested. In fact, although it has been applied to fit the mass-radius relation of white dwarfs, and the emission of gravitational waves from the coalescence of binary black holes \citep{Moffat2016,lopez-romero,Banerjee2017}, its black hole solutions \citep{moffat_blackhole1} have not yet been sufficiently tested.  \citet{Perez2017} investigated the impact on  a specially constructed radiative model and the corresponding spectral energy distributions of a thin accretion disk for both Schwarzschild and Kerr black holes in STVG, without finding substantial differences with observations. Recently, \citet{rdm} investigated the strong field-regime of a rotating black hole solution in STVG studying how the shadow changes with respect to GR. They found that, as the parameter $\alpha$ increases, the modified geometry of the event horizon and of the ergosphere results in generally larger and less asymmetric (more circular) shadows, pointing out that horizon-scale observations of such compact objects could in principle detect signatures of STVG. Thus, our analysis offer a complementary study using the orbital motion of  S2-star around 
the SMBH in the centre of the Milky Way.
 
Our analysis is based on publicly available astrometric and spectroscopic data that have been collected during the past thirty years. Specifically, we used 145 astrometric measurements of the positions and 44 spectroscopic measurements of the radial velocity from \citet{gillessen}, and the measurement of the orbital precession from \citet{gravity}. Our model is based on the Schwarzschild-like solution in STVG outlined in Eq. \eqref{eq:stvg-sch-metric}, and the numerical integration of the corresponding geodesic equations obtained by means of Eq. \eqref{eq:geodesic-equations}. Finally, the 15 dimensional parameter space reported in Eq. \eqref{eq:mcmc_parameters} was explored using the MCMC sampler built in    \texttt{emcee}  \citep{emcee}. We have carried out two analysis with two different approaches:  CA represents our conservative approach  where, since  we expect only small deviations from GR, we adopted Gaussian priors centered on the best fit parameters provided by previous analyses in  \citet{Plewa} and  \citet{gravity}, as listed  in second  and third  columns of Table \ref{tab:priors}; on the other hand, QA represents an approach where, due to the lack of {\em a priori} knowledge of this gravitational system in STVG, we have set flat priors on all model parameters,  except for the orbital period $T$ and pericentre passage $t_p$, as reported in the fifth column of Table \ref{tab:priors}. In both approaches, we did not find any deviation from GR. In Table \ref{tab:results}, we have reported the median values and the 68\% confidence intervals as computed from the posterior distributions  of our MCMC runs. It stands out that our constrained parameter are compatible with one from \citet{gillessen} and \citet{gravity} at most at $3\sigma$.
Finally, and most importantly, we constrained the mean value of the STVG parameter, $\alpha$, to be  $0.027$ and $0.041$, with an upper limit at 99,7$\%$ confidence level of $\lesssim 0.410$ and $\lesssim 0.548$ for the CA and QA approaches,  respectively. This represents the first constraint on 
the dimensionless parameter $\alpha$ which, at those scale of masses, appears in the 
gravitational coupling as $G=G_N(1+\alpha)$ \citep{moffat_blackhole1, Moffat_2021}. Finally, this analysis will also serve as a pilot and reference study for future ones as other relativistic effects on the orbit of S2, in addition to the orbital precession, will be measured.

\section*{Acknowledgements}
IDM acknowledges support from MICINN (Spain) under de project IJCI2018-036198-I. IDM is also supported by Junta de Castilla y León (SA096P20), and Spanish Ministerio de Ciencia, Innovación y Universidades and FEDER (PGC2018-096038-B-I00).  MDL acknowledges INFN Sez. di Napoli (Iniziative Specifica TEONGRAV and QGSKY).

\section*{Data Availability}

Data used in this article are publicly available on the electronic version of \citet{gillessen} at \url{https://iopscience.iop.org/article/10.3847/1538-4357/aa5c41/meta\#apjaa5c41t5}.



\bibliographystyle{mnras}
\bibliography{biblio} 



\bsp	
\label{lastpage}
\end{document}